# The fine band gap modulation effects of aGNRs by organic functional group: A first-principles study

Nuo Liu[1,2,5], Zheqi Zheng[1], Yongxin Yao[2], Guiping Zhang[3,2], Ning Lu[4], Pingjian Li[1], Caizhuang Wang[2] and Kaiming Ho[2]

[1] School of Microelectronics and Solid State Electronics, University of Electronic Science and Technology of China, Chengdu 610054, P. R. China.

[2] Ames Laboratory- U.S. Department of Energy and Department of Physics, Iowa State University, Ames, Iowa 50011, U.S.A.

[3] Department of Physics, Renmin University of China, Beijing 100872, P. R. China

[4] IL Motorola Solutions Inc., IL, 60196, U.S.A.

[5] Author to whom any correspondence should be addressed.

E-mail: liunuo2002@gmail.com

## ABSTRACT

We report a first-principles study of the electronic structure of functionalized graphene nano-ribbon (aGNRs-f) by organic functional group ($CH_2C_6H_5$) and find that $CH_2C_6H_5$ functionalized group does not produce any electronic states in the gap and the band gap is direct. By changing both the density of the organic functional group and the width of the aGNRs-f, a band gap tuning exhibits a fine three family behavior through the side effect. Meanwhile, the carriers at conduction band minimum and valence band maximum are located in both $CH_2C_6H_5$ and aGNR regions when the density of the $CH_2C_6H_5$ is big; while they distribute dominantly in aGNR conversely. The band gap modulation effects make the aGNRs-f good candidates with high quantum efficiency and much more wavelength choices range from 750 to 93924 nm both for lasers, light emitting diodes and photo detectors due to the direct band gap and small carrier effective masses.

# INTRODUCTION

In bottom-up approach, the rational design and synthesis of nanoscale materials will benefit significantly from the work towards understanding fundamental properties and predicting the key structural, chemical, and physical properties. Since the first experimental measurement of graphene in 2004[1], it has become one of the hottest materials because it is the world's thinnest, strongest, and stiffest material, as well as its unique electronic band structure such as massless Dirac fermion physics[2-4] and a half-integer quantum Hall effect[5-7] at room temperature, extraordinarily high carrier mobility, and high thermal conductivity. Extensive research had been conducted to estimate the effects of contacts on the electronic transport through graphene-based material [8]. Though well describes the low energy spectrum and zero band gap of infinite graphene and three-family behavior of band gap in armchair graphene nanoribbon (aGNR), tight binding approximation predicted zero band gap of some kind of aGNRs in contrast with first principles calculation [9]. Therefore, systematic and accurate predictions of the electronic and transport properties of graphene from first principles are an essential topic for applications of future graphene-based devices.

On the application side, aGNRs are excellent candidates for lasing gain emitter since they are found to be stable. Moreover, graphene based photo detectors (PDs) can work over a more wavelength range, while it can have a faster response compared to traditional PDs. Thus, the first-principles calculation of the optical excitations of the edge H and F modified aGNRs were carried out using a many-body perturbation theory approach based on a three-step procedure[10-11] and exact diagonalization of the Hubbard model [12] However, from the point of view of understanding fundamental properties, the use of aGNRs for applications in photo electronics suffers from a major drawback: the three-family behavior of band gaps in aGNRs do not satisfy the further requirements over a much more wavelengths range in electronic devices, lasers, light emitting diodes (LEDs), PDs and optical communications. Therefore, band gap engineering of aGNRs is very important for photo electronic applications.

The functionalization of aGNRs is a promising way to modulate a band gap. The edges hold a great potential for various modifications because of the dangling bonds there. Many researchers have been concentrated on tunable electronic properties of graphene and GNRs including the decoration with organic and inorganic atoms and molecules, and chemical modification of the large graphene surface and edge by covalent, noncovalent interaction by H, F, N, B, P, O, and $NO_2$, $N_2O_4$, NO, $NH_2$ and OH [13-25]. On the one hand, it is expected to obtain desirable properties and get semiconductivity with a wide range of band gaps [26] since they offer new properties that could be combined with the properties of graphene such as conductivity when organic molecule's extended aromatic character is perturbed. On the other hand, importantly, it was shown the strong covalent binding of organic molecules on semiconductor surfaces (or edges) makes those surfaces (or edges) ideal for the immobilization of functional organic materials. This property is useful in the development of new semiconductor-based hybrid materials, sensors, electronic devices, photoelectronic devices, as inorganic semiconductor surfaces (or edges) are certainly absent of various functions of organic materials and those functions and properties can be tuned finely because of the availability of a myriad of organic molecules [27-40]. Obviously, integrating functions of organic materials into inorganic semiconductor-based devices has opened a very promising door for a wide spectrum of technological applications.

Recently, some covalent organic functional group such as tetracyanoethylene (TCNE) [35] and enzyme[36] of aGNRs (armchair) or graphene were studied. Their conductivity is controllable since the molecules act as electronic donors or acceptors. It is reported the adsorbed toluene ($CH_3C_6H_5$) on graphene[37] is a donor. Nevertheless, what about the functional effect on the electronic properties at the edge of aGNR by covalent interaction? Can it exhibit new features different from the adsorbed $CH_3C_6H_5$ on graphene? Motivated by the valuable application, we studied the organic functional group ($CH_2C_6H_5$) effect on the electronic structures of aGNRs (aGNRs-f) by first principles methods in order to better understand the electronic structures and quantum confinement effect (QCE) in these aGNRs-f. The fine three-family band gap tuning effects and charge distribution at high density of the functional group are discussed, which contributes to the fundamental organic/inorganic physics and photoelectronics in the future.

Methods and structures

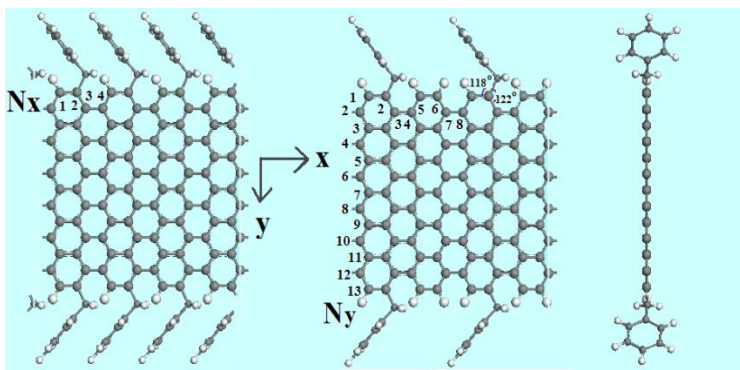

Fig. 1. The structures of the top view of the 13-aGNR-f with Nx being=4 (a) and 8 (b), respectively and the side view (c). There are four unit cells for the former and two for the latter.

Our DFT calculations were performed with the VASP[41] plane-wave-basis codes based on the projector augmented wave (PAW) method. The generalized gradient approximation (GGA) for the exchange-correlation energy functional of the Perdew-Burke-Ernzerhof (PBE) form[42] is adopted. The plane wave cutoff energy is set to 600 eV. The number of centered Monkhorst-Pack **k** points was fourteen along x axial direction with Methfessel-Paxton smearing method in the total-energy calculations. The atomic positions and lattice parameters were fully optimized using the conjugate gradient method. The calculations were converged to the order of $10^{-4}$ eV per cell.

Fig. 1 shows the structures of the energetically most favorable covalent functionalized configuration at side for 13-aGNRs-f with $N_x$ being 4 (Fig.1(a)) and 8 (Fig.1(b)). The aGNRs-f were constructed by periodically repeating the unit cell along x axial direction, and a large vacuum layer with width of 12 Å was used in the y and z directions in order to prevent any spurious interactions between the periodic images. Tests have been performed to make sure that all of the results were converged with respect to the energy cutoff, k-point sampling and vacuum spacing. The dangling bonds on the edge of aGNR were saturated with hydrogen atoms. The period of covalent edged $CH_2C_6H_5$ organic functional group along x direction is labeled as $N_x$ (see Fig.1). The $C_6H_5$ plane of the $CH_2C_6H_5$ is perpendicular to that of the aGNR. The bond length of the C-C between the edged $sp^2$ C atom and that at the end of the $CH_2C_6H_5$ bonds is 1.54 Å which is bigger than that of the $sp^2$ C-C (1.42 Å), and the two associated angles change to 118° and 122°, respectively, which indicate these C-C bond angles deviate from the standard 120° angle of $sp^2$ hybridization and the bonding configurations of $sp^3$ hybridization are formed at the edge.

## Results and discussion

Fig. 2 shows the band structures with $N_x = 4$ and 8. For comparison, the band structures of three pure aGNRs are also shown. Interestingly, the aGNRs-f is also a semiconductor as the fermi level located in the gap. All of the aGNRs-f show a direct gap analogy to aGNRs at the $\Gamma$ point, the gap values as well as the k dispersion along the Γ–X direction strongly depends on the width of the aGNR-f. The $CH_2C_6H_5$ in aGNRs-f do not produce any electronic states in the band gap. For the 9-aGNR-f, 11-aGNR-f and 13-aGNR-f with $N_x$ being 4, the band gap energy ($E_g$) shows oscillatory characteristics with increasing width $N_y$ of the aGNR-f. For 7-aGNRs-f with $N_x$ being 12, it is noted there is a separated energy band with width of 572.7 meV below Fermi level and it does not overlap with the other band, including the conduction band minimums (CBM). PDOS show that they come from the hybridization of C atoms on aGNR and $CH_2C_6H_5$. With $N_x$ increasing to 16, the similar cases appear. Importantly, the two bands below and above Fermi levels separate from the other bands completely in energy and become the band gap states in the gap, which is not desirable for electronic devices. Obviously, the effective band gap modulated function without band gap states of $CH_2C_6H_5$ can only work just for $N_x$ of 4 and 8 of which the density of the organic functional group is big enough.

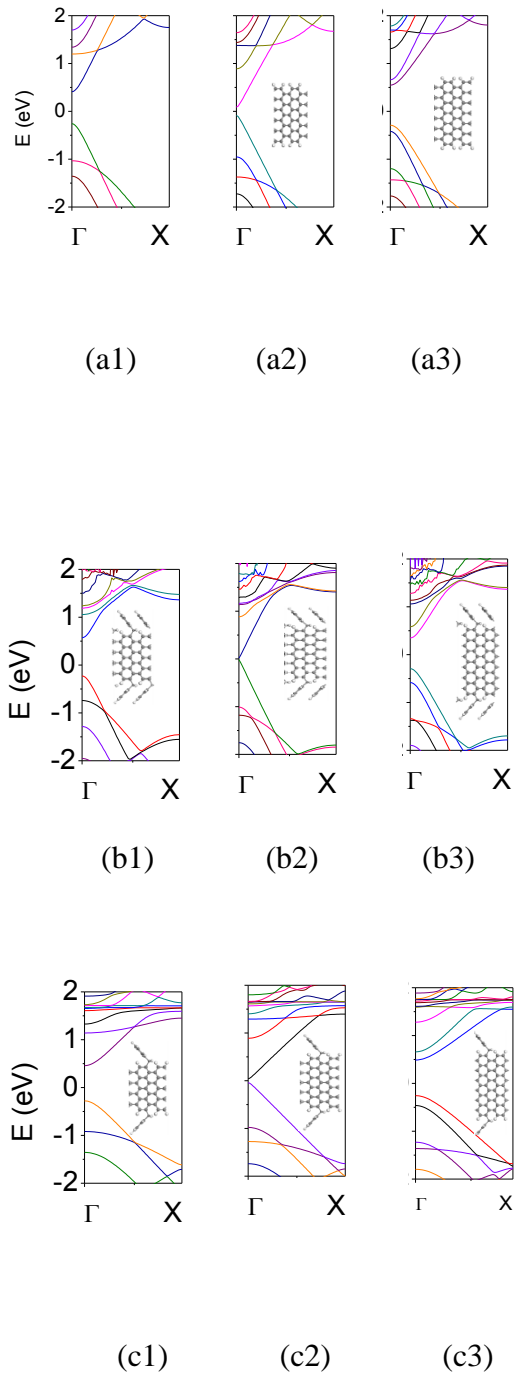

(a1)    (a2)    (a3)

(b1)    (b2)    (b3)

(c1)    (c2)    (c3)

Fig. 2. The band structures of (a) 9, 11 and 13-aGNRs; (b) 9, 11 and 13-aGNRs-f with $N_x=4$; (c) 9, 11 and 13-aGNRs-f with $N_x=8$, respectively. The inserts are the top view of structures of the aGNRs.

As shown in Fig. 3(a), the band gap of a set of aGNR-f also shows the oscillatory decrease of band gaps with the increase of aGNR-f width which is similar to the aGNR. The oscillatory band gaps for the aGNRs were explained by Fermi wavelength in the direction normal to the aGNRs direction. We note that the $CH_2C_6H_5$ at the edges of the aGNR-f modify the electronic structure in the gap vicinity. It is different from the adsorbed toluene ($CH_3C_6H_5$), which is a donor, on graphene[37].

The band gaps of the aGNRs are also divided into three family, with $N_y =6m+1$, $6m-1$, and $6m+3$ (m=1, 2, 3), except for the case of $N_y =3$. The $N_y$ is the same as Barone et. al.'s [9, 43] three-family results although it looks different from their report since the width of the studied aGNRs-f in the work is odd number. For the aGNRs-f of $N_y =6m+1$ and $N_y =6m-1$ (m=1, 2, 3) for aGNR-f with the same width, there is $E_g$(aGNRs-f with $N_x$=4)<$E_g$ (aGNRs-f with $N_x$=8)<$E_g$ (aGNR). In the meanwhile the small band gap appears for the aGNR-f of $N_y =6m-1$ with $N_x$=4, and we have the smallest gap of 0.0132 eV for the 5-aGNR. Thus, the aGNR-f of $N_y =6m-1$ make the gap become smallest when $N_x$=4. On the contrast, it shows that $E_g$ (aGNRs-f with $N_x$=4)>$E_g$ (aGNRs-f with $N_x$=8)>$E_g$ (aGNR) when $N_y =6m+3$(m=1, 2, 3). In the presence of $CH_2C_6H_5$, the gaps remain direct at $\Gamma$ point, but expands to 0.9975 (1.5 times of $E_g$ (aGNR)), 0.5550 (1.2 times of $E_g$ (aGNR)) and 0.4426 eV (1.4 times of $E_g$ (aGNR)) for $N_y$ is 9, 15 and 21, respectively with $N_x$ being 4. A similar effect is obtained with gaps of 0.7358, 0.4884 and 0.3709 eV for $N_y$ is 9, 15 and 21, respectively with $N_x$ being 8. These results, obtained from density of $CH_2C_6H_5$ is 4 and 8, show a strong band gap dependence on $N_x$.

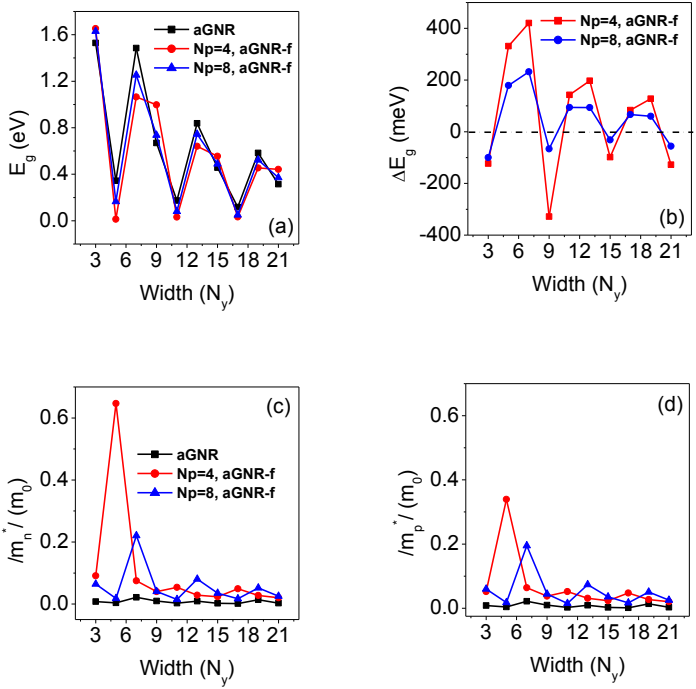

Fig. 3. Calculated (a) band gap energy, (b) variation of the band gap energy, (c) electron effective masses and (d) Hole effective masses as a function of width for hydrogen saturated aGNR (black) and aGNR-f (red and blue).

The band gap difference ($\Delta E_g$, see Fig. 3 (b)) is defined as $\Delta E_g = E_g(aGNR)-E_g(aGNR-f)$, which show the variation of band gap energy of aGNR-f relative to that of the aGNR, in which $\Delta E_g>0$ corresponds to the case of $N_y=6m\pm1$ and $\Delta E_g<0$ for $N_y=6m+3$. The biggest $\Delta E_g$ is about +420 meV for 7-aGNR-f with $N_x=4$ and the smallest one is about -328 meV for 9-aGNR-f with $N_x=4$. Moreover, the band gap modulation effect of $E_g$ ($N_x=4$)> $E_g$ (aGNR) when $N_y=6m+3$ still works with m=4, 5 and 6. Therefore, the functionalized group can shrink the band gap when $N_y=6m\pm1$ and expand it when $N_y=6m+3$. It is obvious the absolute value of $\Delta E_g$ of aGNRs-f with $N_x=4$ is remarkably bigger than that with $N_x=8$. Thus, the fine band gap modulation effects exist for the high density of functionalized group with $N_x=4$ and 8, especially that with $N_x=4$ since both of the biggest and smallest band gap energy appear in them. This implies that the semiconductor behavior of the aGNR-f could be controlled by the density of functionalized group, say, $CH_2C_6H_5$ can significantly modulate the gap finely. Interestingly, the $\Delta E_g$ of fully hydroxylized 7 and 8-aGNRs is -0.7 and +0.2 eV, respectively[22] and that of the 7 and 8-aGNRs-f are -0.4 and +0.32 eV, indicating the similar band gap energy modulation of functionalization. The $\Delta E_g$ of 8-aGNR-OH is bigger than 8-aGNR-f and that of 7-aGNR-OH is smaller than 7-aGNR-f. Obviously, the difference of $\Delta E_g$ between them come from the structural difference. According to Table 1, the threshold wavelength decided by the band gap of the aGNRs-f and aGNRs show

Table 1. The threshold wavelength of the a-GNRs-f and aGNR

| | $N_y$ | 3 | 5 | 7 | 9 | 11 | 13 | 15 | 17 | 19 | 21 |
|---|---|---|---|---|---|---|---|---|---|---|---|
| aGNR | | 811 | 3602 | 835 | 1852 | 7228 | 1479 | 2715 | 10615 | 2127 | 3936 |
| a-GNRs-f | $N_x=4$ | 750 | 93924 | 1165 | 1243 | 39358 | 1934 | 2234 | 37343 | 2725 | 2801 |
| | $N_x=8$ | 761 | 7500 | 990 | 1685 | 15459 | 1666 | 2538 | 24747 | 2371 | 3343 |

The threshold wavelength (nm)

the wavelengths range from 750 to 93924 nm, and 811 to 10615 nm, respectively. Obviously, the aGNRs-f cover a much more wavelengths (λ) from red to far infrared (IR) than that of aGNRs (IR) in photoelectronic devices and allows the different working wavelengths to be controlled finely by both the density of functionalized group and the width of aGNRs-f.

It is well known that mobility is one of the most important parameters of charge transportation, in which carriers effective mass ($m^*$) determine the mobility of carriers according to the formula $\mu=q\tau/m^*$ (The mean free time ($\tau$) is another factor). It is well known $m^*$ near the CBM and VBM is determined by $m^*=h^2(dE^2/dk^2)^{-1}$. Fig. 3 (c) and (d) shows the $m^*$ of the aGNR-f with $N_x$ being 4 and 8. There are four characteristics:(1) The $|m_p^*|$ and $|m_n^*|$ of aGRN are smaller than that of aGNR-f, indicating the carriers of aGNR-f have lower mobility than aGNR ; (2) The biggest m* exists in the structures that $N_y$ is 5 for aGNR-f with $N_x$ being 4, and $N_y$ is 7 with $N_x$ being 8, suggesting modulation brings in heavier effective mass ; (3) An anomalous effective masses appear for $N_y$ is 3 and 5 that $|m_p^*|<|m_n^*|$ (eg, $|m_p^*|=0.3393$ and $|m_n^*|=0.6466$) with $N_x=4$; and finally, we have $|m^*|$ ($N_x=8$)<$|m^*|$($N_x=4$) for electron (hole) when $N_y$ is 3 (only for $|m_n^*|$), 5, 11and 17, respectively. In short, carriers effective masses are controlled not only by width of aGNR but also by the density of the edged functional group.

In order to study the functionalization of $CH_2C_6H_5$ on the aGNR-f, the TDOS of the aGNR-f (black line) and the DOS (red line) for the edged $CH_2C_6H_5$ respect to aGNR (blue line) are shown in Fig. 4. We focus on the electronic states in the vicinity of the Fermi level because they are mainly responsible for the carrier's transition. The band gap seems to disappear for the DOS of 5, 11 and 17-aGNR-f near Fermi level since the width of their band gaps are very small (e.g., $E_g$ =0.0132 eV for 5-aGNR-f) and the expansion of the DOSs below and above Fermi level lead to their merges. The DOS from $CH_2C_6H_5$ do not introduce any band gap states in the gap, and locate away from their valence band maximum (VBM) and the conduction band minimums (CBM) at the Γ point (the center of the Brillouin zone) for the aGNRs-f with $N_x$=4 and 8, which is different from the doping in GNR and carbon nanotubes [44-48] that cause the formation of electronic states within the gap. The width of the peak around 1.5 eV of aGNR-f with $N_x$ =8 is narrower than that of aGNR-f with $N_x$=4, because the density of the $CH_2C_6H_5$ in a unit cell of the latter is two times of the former and make the peak expanse. For the aGNR-f with $N_x$=4, the partial DOS (PDOS) shows that the p orbital electrons from the four C atoms from the benzene ring on the two sides of the symmetric axis of $CH_2C_6H_5$ decide the PDOS at low energy in conduction band and all of the p orbital electrons from the seven C atoms of $CH_2C_6H_5$ decide that at high energy in valence band. However, it is noted that, for 3-aGNR-f and 5-aGNR-f with $N_x$=4, the $CH_2C_6H_5$ at the edges of the aGNRs strongly modify the electronic structure in the gap vicinity, there being strong hybridization in between the the p orbital electrons from the C atoms of aGNR and that of $CH_2C_6H_5$ near CBM. Moreover, there is strong hybridization of PDOS between the two C atoms at VBM for 5-aGNR-f with $N_x$=4 in energy, leading to a bigger band gap. We note that the maximum of our gap variation is about 50%, indicating the strong band gap modulation effects for aGNRs-f with different density of functionalization and width.

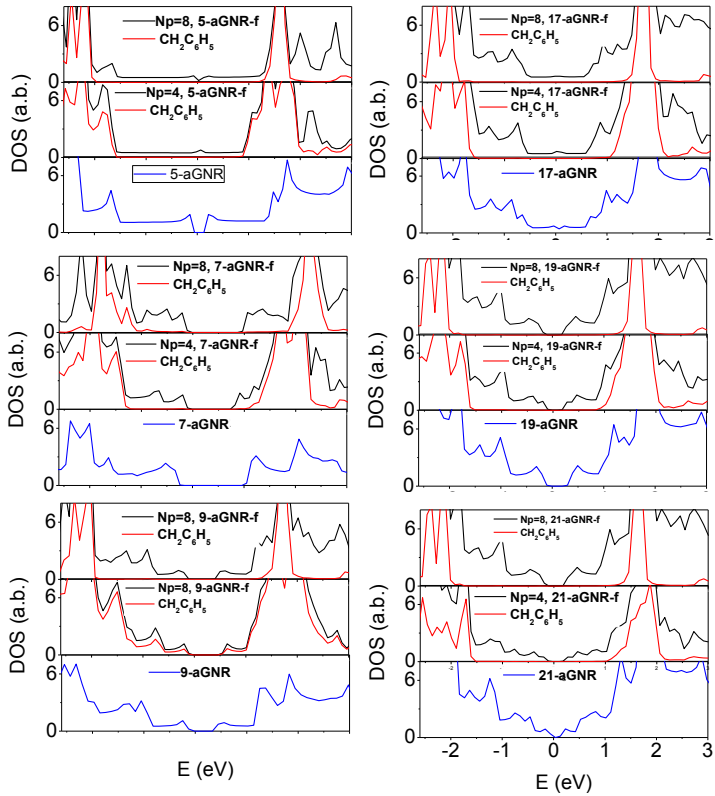

Fig. 4. The densities of states corresponding to the 5, 7, 9, 17, 19 and 21-aGNRs (blue) and -aGNRs-f (black), and the densities of states corresponding to the functionalized group (red) by using Gaussian broadening of 0.05 eV.

When $N_x$ is 4, the low energy of the peak of $CH_2C_6H_5$ near CBM is closer to CBM than that of $N_x$ is 8. Considering the peak width of the former is wider than that of the latter, we reckon this implies stronger interaction between $CH_2C_6H_5$ and the C atoms in the aGNR-f. The same results are also seen in valence band, say, when $N_x$ is 4, the high energy of the peak of $CH_2C_6H_5$ near VBM is closer to VBM than that of $N_x$ is 8. Therefore, it suggests the modulation function of high density functional group is more remarkable. Our results of functionlization of $CH_2C_6H_5$ are consistent with the NH$_2$ functionlization, B, N edge substitutions 12-aGNRs [49], both of which have the sp$^3$ configuration at the edge of aGNRs. However, it is in contrast with that physisorbed NH$_3$ group[46] and TCNE[35], substitutionally doped with a single B, P and N [50,51] in aGNR which show the site and spin-dependence acceptorlike or donorlike states in the band gap.

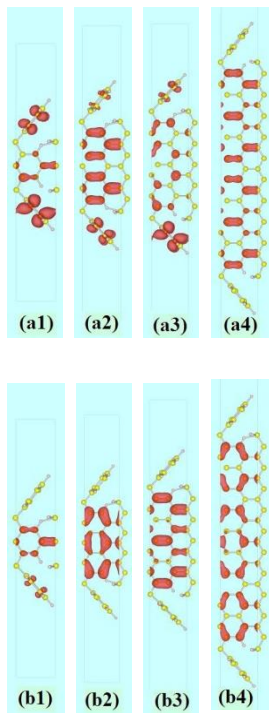

Fig. 5. Top view of the distribution of charge density (red) plots at the isovalue of 0.003e of 3, 7, 9 and 19–aGNRs-f of a unit cell for the bottom of conduction band (up) and top of valence band (down), respectively. Yellow balls are C atoms and pink small balls are H atoms, respectively.

To gain more insights into the nature of the electronic states at the CBM and VBM, Fig. 5 presents the distribution of charge density corresponding to the electronic states at CBM and VBM in real-space. The charge densities are plotted at an iso-value of 0.003e on the top views of the aGNR-f. The charges at the CBM in the aGNR-f with $N_x$ being 4 are distributed in both $CH_2C_6H_5$ and GNR region when $N_y$ is small. It shows the electrons strongly localized on the four

C atoms from the benzene ring on the two sides of the symmetric axis in the centers of $CH_2C_6H_5$, which is in agreement with the PDOS, and the $sp^2$-bonded C atoms of aGNR in CBM. With width increasing, the charges distributed in $CH_2C_6H_5$ decrease and that in GNR region increase. It is remarkable the charge distribution on $CH_2C_6H_5$ in 9-aGNR-f is bigger than that of 7-aGNR-f, corresponding to the DOS in Fig.5. Therefore, it demonstrates the strong influence of $CH_2C_6H_5$ further in aGNR-f with $N_y$ is 9. Moreover, the charges are located on the GNR region completely when the width is bigger enough. The same phenomenon is also observed in aGNR-f with $N_x$ being 8. Obviously, the influence of the $CH_2C_6H_5$ decreases with the increasing of the number of C atoms. It suggests the strong hybridization at CBM resulting from the two different C atoms on GNR and $CH_2C_6H_5$ when the width is small. Furthermore, the electronic states of VBM are confined in the GNR region since the electron cloud of VBM is shared by C atoms in the GNR region (see Fig. 5 (a)). On the other hand, the electronic states of CBM are mainly decided by $CH_2C_6H_5$ and partly by GNR. The electron cloud of VBM is shared mainly in the GNR region (Fig. 5 (b)) and partly in $CH_2C_6H_5$ region for 3-aGNR-f. With the width increasing, the charges localized in $CH_2C_6H_5$ region disappear, and they distribute near the C atoms of aGNR, indicating the electrons strongly localized on C atoms of aGNR in VBM. Thus, the effect of modulated function of the $CH_2C_6H_5$ on VBM can be neglected with $N_y$ increasing. Similarly, for the aGNR-f with $N_x$ being 8, the analogue charge distribution is also observed. The charge distribution in $CH_2C_6H_5$ of the aGNR is smaller than that with $N_y$ is the same, suggesting the modulated function of $CH_2C_6H_5$ decrease due to the increasing of the C atoms in aGNR region when the width is the same which is also showed in Fig. 3(b) that $\Delta E_g$ of the aGNR-f with small $N_x$ is bigger than that with big $N_x$. Therefore, the spatially charges distribution at CBM and VBM strongly suggest the controlling function of the $CH_2C_6H_5$ functional group is in agreement with the variation of the band gap energy in Fig. 3(b), and both the quantum confinement effect due to size effect and the band gap size modulation effect originating from $CH_2C_6H_5$ functional group can be realized in the aGNR-f along x direction.

The main value of the a-GNR-f is the fine band gap modulation effects of the $CH_2C_6H_5$ functional group. In other words, there are strong quantum confinement effects on the electronic states on aGNR-f. The effects are valuable for photoelectronic devices in much more wavelengths from 750 to 93924 nm, by changing the density of the functionalized group and the width of the aGNRs-f. There is not any phonon participating the generation and recombination of carriers due to the direct band gap, leading to the higher quantum efficiency of photoelectronic devices. However, the carriers recombination like that in traditional p-n junction devices still exist in aGNR–f and aGNR because the electrons and holes are not separated in space, suggesting the decreasing of the quantum efficiency in solar cells, which can be improved by rational design of the structure of the devices. The other merit application is for the high electron mobility like high electron mobility transistors (HEMTs). With small carrier effective masses, the high mobility is realized in the aGNR–f and aGNR. The work may help to promote the integration functions of organic materials into inorganic semiconductor-based photoelectronic and electronic devices in the future.

Summary

In conclusion, we have studied the band gap modulation effect of organic functional group based on first-principles calculations. Our calculations indicate that the $CH_2C_6H_5$ functionalized group does not give any electronic states that pin the Fermi energy in the gap when the density of functionalized group is 4 and 8, respectively. There is fine three-family band gap tuning resulting from the side-effect of the functional group since the direct band gaps can be tailored by both the density of the organic functional group ($N_x$=4,8) and the width of aGNR-f. The carriers's effective masses are also modulated by the two same factors. Detailed investigations reveal that the electrons localized on both the four C atoms from the benzene ring on the two sides of the symmetric axis in the center of CH2C6H5 and the C atoms of aGNR in CBM when the width is small. With the width and density of the organic functional group decreasing, respectively, the charges distributed in $CH_2C_6H_5$ decrease and that in GNR region increase. Finally, they migrate to the GNR region completely, suggests the decreasing modulated function of the organic functional group. The analogy charge distribution also exists in VBM. The edged covalent bonding in aGNRs could be a good method to modulate the electronic property of aGNR because it modulates the band gap without destroying the linear energy dispersion near the Dirac point. The applications of a-GNRS-f as high quantum efficiency and photoelectronic devices over much more wavelengths (710-93924 nm) and high mobility transistors are discussed.

## Acknowledgments

Ames Laboratory is operated for the U.S. Department of Energy by Iowa State University under Contract No. DE-AC02-07CH11358. This work was supported by the Director for Energy Research, Office of Basic Energy Sciences including a grant of computer time at national energy research Supercomputing center (NERSC) in Berkeley. Nuo Liu's work at Ames Laboratory was supported by the International Corporation and Communication Scholarship of Sichuan Province (Grant No. 2012HH0027) and the National Natural Science Foundation of China (Grant No. 51202022). G. P. Zhang acknowledged the support by the National Natural Science Foundation of China (Grant No. 11204372).